       \newcommand{\beq}{\begin{equation}}
       \newcommand{\eeq}{\end{equation}}
       \newcommand{\beqa}{\begin{eqnarray}}
       \newcommand{\eeqa}{\end{eqnarray}}
       \newcommand{\beqas}{\begin{eqnarray*}}
       \newcommand{\eeqas}{\end{eqnarray*}}
       
       \newcommand{\nab}{\nabla}
       \newcommand{\bnab}{\mbox{\boldmath ${\nabla}$}}

       \newcommand{\x}{\mbox{\boldmath$\times$}}

       \newcommand{\bb}{{\mathbf b}}

       \newcommand{\bn}{{\mathbf n}}

       \newcommand{\bv}{{\mathbf v}}
       
       \newcommand{\bx}{{\mathbf x}}

       \newcommand{\bA}{{\mathbf A}}
       \newcommand{\bB}{{\mathbf B}}

       \newcommand{\bJ}{{\mathbf J}}

       \newcommand{\bM}{{\mathbf M}}
       \newcommand{\bN}{{\mathbf N}}

       \newcommand{\bV}{{\mathbf V}}

       \newcommand{\bxi}{\mbox{\boldmath $\xi$}}

\newcommand{\nonu}{\nonumber \\}

\documentclass[12pt]{article}
\usepackage[dvips]{graphicx}\usepackage[dvips]{graphics}
\usepackage{enumerate}
\usepackage{psfrag}
 \usepackage{color}
\newcommand{\pars}{\partial}
\begin{document}
\begin{center}
{\large \bf
%On equilibria in Hall magnetohydrodynamics model}
 A sufficient condition for the linear stability
 of magnetohydrodynamic equilibria with  \\\vspace{1mm}field aligned incompressible flows}
\vspace{3mm}

{\large  G. N. Throumoulopoulos}

{\it
University of Ioannina, Association Euratom - Hellenic Republic,\\
%\vspace{-1mm}
 Division of Theoretical Physics, GR 451 10 Ioannina, Greece}
\vspace{2mm}

 {\large H. Tasso}

 {\it  Max-Planck-Institut f\"{u}r Plasmaphysik, Euratom
Association,\\
%\vspace{-1mm}
 D-85748 Garching, Germany }
\end{center}
%\noindent
%
%
%\vspace{2mm}
%\begin{center}
%{\large \it December 2000}
%\end{center}
\vspace{2mm}
\begin{center}
{\bf Abstract}
\end{center}

\noindent A sufficient condition for the linear stability of three
dimensional equilibria with incompressible flows parallel to the
magnetic field is derived. The condition involves  physically
interpretable terms related to the magnetic shear and the flow
shear.
 \vspace{2cm}

%\noindent {\sf To appear in Physics of Plasmas}

\newpage

\begin{center}
{\bf \large I.\  Introduction}
\end{center}

 For static ideal magnetohydrodynamic (MHD) equilibria  there is a powerful tool known as ``the energy
 principle"
 providing necessary and sufficient conditions for linear stability \cite{BeFr}. In the presence of flow, however,
 the stability problem is much tougher because the force operator becomes non Hermitian;
 thus, only sufficient conditions were obtained \cite{FrRo}-\cite{HiYo}.
 Motivation of the present study is a couple of  papers by Ilin and
 Vladimirov \cite{IlVl1,IlVl2}  in which a sufficient condition
 was derived for the linear stability  of plasmas with constant
 density and  incompressible flows parallel to the magnetic field.
 This condition states that an
 %generic
 equilibrium
 %of the above mentioned kind
 is
 stable to three dimensional perturbations provided that: i) the flow
 is  sub-Alfvenic and ii) inequalities (51) of Ref. \cite{IlVl2} are
 satisfied. Here we show, however, that those inequalities are not correct
 %in the presence of flow
  for the
 following  reasons:
 \begin{enumerate}
 \item The authors of Refs. \cite{IlVl1} and \cite{IlVl2}
 have not noticed that because
 of the field aligned flow the equilibrium current density lies on
 magnetic surfaces. This property simplifies the stability analysis and results
 in a single inequality for the sufficient condition
 in place of the couple of inequalities (51) of Ref. \cite{IlVl2}.
 \item A  term associated with the flow  shear
 was ignored in Refs \cite{IlVl1} and \cite{IlVl2}.
 \end{enumerate}
 The correct sufficient
 condition  obtained here contains
 physically interpretable terms related to the magnetic shear and the flow
 shear.

 The equilibrium characteristics are examined  in Sec. II including
 a prove of the coincidence of the current density surfaces with the
 magnetic surfaces. Sec. III reviews the energy principle established
 in Refs. \cite{IlVl1} and \cite{IlVl2} which subsequently is  employed
 in Sec. IV to derive the sufficient condition. A major part of the
 derivation is presented in the Appendix.

 \begin{center}
 {\bf \large II.\ \ Equilibrium}
 \end{center}

 We consider the steady states of a plasma of constant density
 and incompressible flow parallel to the magnetic filed in the framework
 of ideal MHD (see for example Eqs. (1)-(6) of Ref. \cite{TaTh2} written in convenient
 units and the density
 set to unity).
 Also, it is assumed the existence of well defined equilibrium magnetic surfaces in three dimensional
 geometry which are labeled by a smooth function $\psi$.
  Using
 \beq
 \bV = \lambda \bB,
                                                  \label{1}
 \eeq
 where $\lambda$ is an arbitrary function, the incompressibility
 condition ($\bnab \cdot \bV=0$) implies that $\lambda$ is a surface quantity:
 \beq
 \lambda=\lambda(\psi).
                                                   \label{2}
 \eeq
Then, employing the identity $\left(\bV\cdot\bnab\right)\bV =\bnab
V^2/2-\bV \x \bnab\x \bV$, the momentum equation
 $$
 \left(\bV\cdot\bnab\right)\bV=\bJ \x \bB -\bnab P
 $$
 leads to
 \beq
 \left(1-\lambda^2\right)\bJ\x\bB=\bnab\left(P+\frac{\lambda^2
 B^2}{2}\right)-B^2\bnab\left(\frac{\lambda^2}{2}\right),
                                                  \label{4}
 \eeq
 where $B$ is the magnetic field modulus.
 The component of  (\ref{4}) along the magnetic field implies that
 the quantity $P+\lambda^2 B^2/2$ is uniform on magnetic surfaces:
 \beq
 P+\frac{\lambda^2 B^2}{2}\equiv P_s(\psi).
                                                \label{5}
 \eeq
 Thus, owing to the flow the isobaric surfaces depart from the magnetic
 surfaces unlike  the case of static equilibrium associated with the
 surface function $P_s(\psi)$. Consequently, Eq. (\ref{4}) is put in
 the form
 $$
 \left(1-\lambda^2\right)\bJ\x\bB=P_s^\prime\bnab \psi -(\lambda^
 2)^\prime \frac{B^2}{2}\bnab \psi
 $$
 or
 \beq
 \bN\equiv \bJ\x\bB=g(\psi, B^2) \bnab \psi
                                                \label{6}
 \eeq
 where
 \beq
 g(\psi,B^2)\equiv
 \frac{P_s^\prime}{1-\lambda^2}-\frac{(\lambda^2)^\prime}{1-\lambda^2}
 \frac{B^2}{2}.
                                                \label{7}
 \eeq
 %and the prime denotes derivative with respect to $\psi$.
  Eq. (\ref{6}) implies that the current density lies on magnetic surfaces
  a property not noticed in Refs.
  \cite{IlVl1} and \cite{IlVl2}. Note that this holds because of the
   incompressible field aligned flows; for flows of arbitrary direction the current
  surfaces do not coincide with the magnetic surfaces. The fact that
   $\bB$, $\bJ$ and $\bV$ share the same surfaces simplify
   the  stability analysis to follow. To this end we also will  need the quantity
   \beq
   \bM\equiv \bnab\x\bN=\bnab g\x\bnab\psi
                                                   \label{7a}
   \eeq
   from which it follows that
   \beq
   \bM\cdot\bN=0.
                                                    \label{7b}
   \eeq

 \begin{center}
 {\bf \large III.\ \ Review of the energy principle}
 \end{center}

 In Refs. \cite{IlVl1,IlVl2} an energy principle was established  for
 incompressible perturbations [$\bnab \cdot \bxi(\bx,t)$=0] around a steady state
 and  non-slip boundary conditions:
 \beq
  \bv\cdot \bn =\bb\cdot \bn=\bxi\cdot \bn=0.
                                                     \label{7c}
 \eeq
 Here $\bv(\bx,t)$ and $\bb(\bx,t)$ are the perturbations of the velocity and
 the magnetic field and conditions (\ref{7c})
 are imposed  on a fixed boundary $\pars \cal D$ surrounding the plasma
 domain $\cal D$.
  The principle is based on the fact that the perturbation
 energy
 \beq
 E\equiv \int_{\cal
 D}\left(\frac{1}{2}\dot{\bxi}^2-\frac{1}{2}\bxi\cdot\hat{K}\bxi\right)dV,
                                                                \label{8}
 \eeq
 is conserved by the linearized ideal MHD equations ($dE/dt=0$).
 Here $\hat{K}$ is a symmetric operator defined by the formula
 $$
 \hat{K}\bxi=\bV\x\bnab\x\bv+\bv\x\mathbf \Omega-\bB\x\nabla\x\bb-\bb\x\bJ,
 $$
 where
 $\bv=\bnab\x(\bxi\x\bV)$,
 \beq
 \bb=\bnab\x(\bxi\x\bB),
                                                             \label{8a}
 \eeq
 and
 $\mathbf \Omega=\bnab\x\bV$.
 Evidently, $E$ as a quadratic functional of $\dot{\bxi}$ and $\bxi$
 is positive definite if the potential energy
 \beq
 W=-\frac{1}{2}\int_{\cal D} \bxi\cdot\hat{K}\bxi dV
                                                       \label{9}
 \eeq
 is positive definite. It is known, however, that for flows of
 arbitrary direction the functional $W$ is never strictly positive
 definite \cite{FrRo}-\cite{HiYo}. For this reason further consideration is restricted to
 the steady states with field aligned flows described in Sec. II.
 In this case (\ref{9}) can be
 written in the form
 \beq
 W=\frac{1}{2}\int_{\cal D} \left\{
  (1-\lambda^2)\left\lbrack \bb^2+\bb\cdot (\bJ\x \bxi)\right\rbrack
  -2\lambda
 (\bxi\cdot\bnab\lambda)\left\lbrack \bxi\cdot
 (\bB\cdot\bnab)\bB\right\rbrack  \right\} dV.
                                                            \label{10}
 \eeq
 Derivation of (\ref{10}) is given in Ref. \cite{IlVl1}.
 Whenever the potential energy (\ref{10}) is positive definite the
 equilibrium is linearly stable.

 \begin{center}
 {\bf \large IV.\ \ Sufficient condition for linear stability}
 \end{center}

 As in  Refs. \cite{IlVl1} and \cite{IlVl2} assuming that $ \bJ\x\bB\neq
 0$ we express the perturbation vector $\bxi$ in the form
 \beq
 \bxi=\alpha(\bx, t)\bN+\beta(\bx, t)\bJ+\gamma(\bx,t)\bB.
                                                         \label{11}
 \eeq
 It can then be shown (see Appendix) that $W$
 assumes the form
 \beq
 W=W_1+W_2,
                                                           \label{12}
 \eeq
 \beq
 W_1 = \frac{1}{2}\int_{\cal D} (1-\lambda^2)\left(\bb+\alpha \bJ\x\bN\right)^2
 dV,
                                                          \label{13}
 \eeq
 \beq
 W_2=\int_{\cal D} A \alpha^2,
                                                        \label{14}
 \eeq
 where
 \beq
 A=-(1-\lambda^2)\left(\bJ\x\bN\right)\cdot(\bB\cdot\bnab)\bN
   -\lambda(\bN\cdot\bnab\lambda)\left(\bN\cdot\frac{\bnab B^2}{2}+N^2\right).
                                                        \label{15}
 \eeq
 Evidently, $W$ is positive semidefinite if $\left|\lambda\right|\leq 1$ and
 \beq
 A\geq 0 \ \ \mbox{in}\ \ {\cal D}.
                                                       \label{16}
 \eeq
 Inequality (\ref{16}) is substantially different from the
 respective inequalities (51) of Ref. \cite{IlVl2}. In particular,  the
 last term of (\ref{15}) containing $\bnab \lambda$ was missed in
 \cite{IlVl1} and \cite{IlVl2}.
 Using the equilibrium relations of Sec. III,
 (\ref{15}) reduces to
 \beqa
 A&=&-g^2\left\{(1-\lambda)^2\left(\bJ\x\bnab
 \psi\right)
 \cdot(\bB\cdot\bnab)\bnab\psi \right. \nonu
 & & \left. -\frac{(\lambda^2)^\prime}{2}\left|\bnab
 \psi\right|^2\left(\bnab \psi\cdot\frac{\bnab B^2}{2}+g|\bnab
 \psi|^2\right)\right\}.
                                                      \label{17}
 \eeqa
 On account of (\ref{12})-(\ref{14}) and (\ref{17}) we can conclude
 that  {\em a general steady state of a plasma of constant density and incompressible
 flows parallel to the magnetic is stable to small three-dimensional
 perturbations if  i) the flow is sub-Alfv\'enic and ii)
 \beq
 \tilde{A}\equiv \frac{A}{g^2} \geq 0.
                                                       \label{18}
 \eeq }
 Using the relation
 $$
 (\bB\cdot\bnab)\bnab \psi=\bJ\x\bnab \psi
 -(\bnab\psi\cdot\bnab)\bB
 $$
 $\tilde{A}$ can be put in the physically interpretable form:
 \beqa
 \tilde{A}&=&-(1-\lambda)^2\left\lbrack (\bJ\x\bnab
 \psi)^2-(\bJ\x\bnab\psi)\cdot(\bnab \psi\cdot\bnab)\bB\right\rbrack
 \nonu
 & & -\frac{(\lambda^2)^\prime}{2}|\bnab\psi|^2\left(\bnab\psi\cdot\frac{\bnab B^2}{2}+g|\bnab
 \psi|^2\right).
                                                    \label{19}
 \eeqa
 The first negative destabilizing term in (\ref{19}) should be
 related to current driven modes. The other terms can be either
 stabilizing or destabilizing. This depends on the sign of $(\lambda^2)^\prime$
 in relation to the velocity shear and on
 the differential variation of  $\bB$ and $B^2$ perpendicular
 to the magnetic surfaces in  relation to the magnetic shear.
 Also, the last term has an additional
 implicit dependence on $(\lambda^2)^\prime$ and $P_s^\prime$
 through the quantity $g$ [Eq. (\ref{7})].

 It is recalled that the sufficient condition established here can
 be applied to any steady state without  geometrical restriction.
 Application to steady states of fusion concern in connection with  possible
 stabilizing  effects of the flow is under way.

 \begin{center}
 {\large \bf Appendix:\ \ Derivation of (\ref{12})-(\ref{15})}
 \end{center}

 The procedure to follow is based on that of Appendix of Ref.
 \cite{IlVl2}. Since there are substantial differences, however,
 the derivation will be presented in a self contained way.

 Preliminarily, in view of the representation (\ref{11}) for $\bxi$
 and the
 incompressibility condition $\nab \cdot \bxi=0$ we obtain  the
 following relations:
 \beq
 \bnab\cdot(\alpha \bN)+\bJ\cdot\bnab \beta+\bB\cdot \bnab \gamma=0,
                                                                    \label{a1}
 \eeq
 $$
 \bxi\x\bB=\alpha\bN\x\bB+\beta \bN, \ \
 \bJ\x\bxi=\alpha\bJ\x\bN+\gamma \bN.
 $$
 Also, (\ref{8a}) becomes
 $$
 \bb=\bnab\x(\alpha \bN\x\bB+b\bN)
 $$
 The first term of (\ref{10})  is written as
 \beqa
 I&\equiv&\frac{1}{2}\int_{\cal D} (1-\lambda^2)\left\lbrack
 \bb^2+\bb\cdot(\bJ\x\bxi)\right\rbrack dV \nonu
 &=&\frac{1}{2}\int_{\cal D} (1-\lambda^2)\left\lbrack
 \bb^2+\bb\cdot(\alpha\bJ\x\bN+\gamma \bN)\right\rbrack dV \nonu
 &=& \frac{1}{2}\int_{\cal D} (1-\lambda^2)\left\lbrack
 \left(\bb+\alpha\bJ\x\bN\right)^2-\alpha^2(\bJ\x\bN)^2\right.\nonu
 & & \left. -\bb\cdot(\alpha\bJ\x\bN)+\bb\cdot(\gamma
 \bN)\right\rbrack dV
                                                                    \label{a2}
 \eeqa
 Employing the identity $\bnab\cdot(\bA\x\bB)=\bA\cdot\bnab\x\bB-\bB\cdot\bnab\x\bA$
 %for any vector functions $\bA$ and $\bB$
 we have in connection with the last term of (\ref{a2})
 \beqas
 \bb\cdot(\gamma
 \bN)&=&\gamma\bN\cdot\bnab\x\left(\alpha\bN\x\bB+\beta\bN\right)
 \nonu
 &=&\left(\alpha\bN\x\bB+\beta\bN\right)\cdot
      \bnab\x(\gamma
      \bN)+\bnab\cdot\left\lbrack(\alpha\bN\x\bB+\beta\bN)\x(\gamma
      \bN)\right\rbrack
 \eeqas
 and therefore
 \beqas
 (1-\lambda^2)\bb\cdot(\gamma\bN)&=&(1-\lambda^2)\left(\alpha\bN\x\bB+\beta\bN\right)\cdot\bnab\x(\gamma\bN)
 \nonu
  & & +
 \bnab\cdot\left\lbrack(1-\lambda^2)\gamma(\alpha\bN\x\bB+\beta\bN)\x \bN \right\rbrack
                                                                     \label{a3}
 \eeqas
 the last term following from the fact that $\bN$ is proportional to
 $\bnab \psi$ [Eq. (\ref{6})].
 Substituting (\ref{a3}) into (\ref{a2}) and integrating by parts
 furnishes
 \beqa
 I&=&\frac{1}{2}\int_{\cal
 D}(1-\lambda^2)\left\lbrack(\bb+\alpha\bJ\x\bN)^2-\alpha^2(\bJ\x\bN)^2\right.
   \nonu [4mm]
   & &\left. \stackrel{Y_2} {\overbrace{-\alpha(\bJ\x\bN)\cdot\bb}} +
      \stackrel{\large Y_1
      }{\overbrace{\bnab\x(\gamma\bN)\cdot\left(\alpha\bN\x\bB+\beta\bN\right)}}\right\rbrack dV.
                                                                    \label{a4}
 \eeqa
 Furthermore, employing (\ref{7a}) and (\ref{7b}) we find for the
 above defined quantities  $Y_1$ and $Y_2$:
 \beqas
 Y_1&=&\bnab\x(\gamma\bN)\cdot\left(\alpha\bN\x\bB+\beta\bN\right)=\left(
        \gamma\bnab\x\bN+\bnab\gamma\x\bN\right)\cdot(\alpha\bN\x\bB+\beta\bN)
        \nonu
    &=& \alpha\gamma\bM\cdot\bN\x\bB-aN^2(\bB\cdot\bnab\gamma)
    %&=&-\alpha^2(\bJ\x\bN)\cdot\x
    \nonu
 %\eeqas
 %\beqas
 Y_2&=&-\alpha(\bJ\x\bN)\cdot\bnab\x\left(\alpha\bN\x\bB+\beta\bN\right)\nonu
    &=&  -\alpha(\bJ\x\bN)\cdot\left\lbrack\bnab\alpha\x(\bN\x\bB)+\alpha\bnab\x(\bN\x\bB)
       +\bnab\beta\x\bN+\beta\bM\right\rbrack \nonu
    &=&-\alpha^2(\bJ\x\bN)\cdot\bnab\x(\bN\x\bB)-\alpha N^2(\bN\cdot\bnab\alpha+\bJ\cdot\bnab\beta)
    -\alpha\beta\bM\cdot(\bJ\x\bN)
 \eeqas
 and with the aid of (\ref{a1})
 $$
 Y_1+Y_2=-\alpha^2\left\lbrack(\bJ\x\bN)\cdot\bnab\x(\bN\x\bB)-N^2\bnab\cdot\bN\right\rbrack
         -\alpha\beta\bJ\cdot(\bN\x\bM)-\alpha\gamma\bB\cdot(\bN\x\bM).
 $$
 Eq. (\ref{a4})  then becomes
 \beqa
 I&=&\frac{1}{2}\int_{\cal D}
     (1-\lambda^2)\left\{(\bb+\alpha\bJ\x\bN)^2  \right. \nonu
     & & -\alpha^2\left\lbrack(\bJ\x\bN)\cdot\bnab\x(\bN\x\bB)
     -N^2\bnab\cdot\bN+(\bJ\x\bN)^2\right\rbrack \nonu
  & & \left. -\alpha
  \beta\bJ\cdot(\bN\x\bM)-\alpha\gamma\bB\cdot(\bN\x\bM)\right\} dV.
                                                           \label{a5}
  \eeqa
  The staff within the square brackets in (\ref{a5}) can be put in
  the concise form
  \beq
  (\bJ\x\bN)\cdot\bnab\x(\bN\x\bB)
     -N^2\bnab\cdot\bN+(\bJ\x\bN)^2=2(\bJ\x\bN)\cdot(\bB\cdot\bnab)\bN.
                                                       \label{a6}
  \eeq
  To show this we employ the relation $\bnab(\bB\cdot\bN)=0$ implying
  that
  $$
  (\bN\cdot\bnab)\bB=-(\bB\cdot\bnab)\bN+\bM\x\bB+\bJ\x\bN;
  $$
  then,
  \beqas
  (\bJ\x\bN)\cdot\bnab\x(\bN\x\bB)&=&(\bJ\x\bN)\cdot
  \left\lbrack
  (\bB\cdot\bnab)\bN-\bB(\bnab\cdot\bN)-(\bN\cdot\bnab)\bB \right\rbrack \nonu
  &
  =&2(\bJ\x\bN)\cdot(\bB\cdot\bnab)\bN-(\bJ\x\bN)^2+N^2\bnab\cdot\bN.
  \eeqas
  We now consider the second part of $W$ [see Eq. (\ref{10})]:
  \beq
  Q\equiv=-\int_{\cal D} \lambda
  (\bxi\cdot\bnab\lambda)\left\lbrack
  \bxi\cdot(\bB\cdot\bnab)\bB\right\rbrack dV.
                                                    \label{a7}
  \eeq
  Using the relations
  $$
  (\bxi\cdot\lambda)=\alpha\bN\cdot\bnab \lambda
  $$
  and
  $$
  \bxi\cdot(\bB\cdot\bnab)\bB=\bxi\cdot\left(\frac{\bnab B^2}{2}+\bJ\x\bB\right)=(\alpha
  \bN+\beta \bJ+\gamma \bB)\cdot\frac{\bnab B^2}{2}+\alpha N^2,
  $$
  (\ref{a7}) is put in the form
  \beqa
  Q&=&-\int_{\cal D}\left\{\lambda\alpha^2 (\bN\cdot\bnab\lambda)
      \left(\bN\cdot\frac{\bnab B^2}{2}+N^2\right ) \right.\nonu
   & & \left. +\lambda \alpha
      \beta(\bN\cdot\bnab\lambda)(\bJ\cdot\frac{\bnab B^2}{2})
      +\lambda \alpha\gamma(\bN\cdot\bnab\lambda)\left(\bB\cdot\frac{\bnab
      B^2}{2}\right)\right\} dV
                                                    \label{a8}
   \eeqa
 The first term in (\ref{a8}) containing $\bnab \lambda$ was ignored in
 Refs. \cite{IlVl1} and \cite{IlVl2}.
 In view of (\ref{a5}), (\ref{a6}) and
 (\ref{a8}),
 %and the fact that $\bJ$ lies on  magnetic surfaces,
 $W$ is written in  the form
 \beq
 W=W_1+W_2,
                                                    \label{a9}
 \eeq
 \beq
 W_1 = \frac{1}{2}\int_{\cal D} (1-\lambda^2) \left(\bb+\alpha \bJ\x\bN\right)^2
 dV,
                                                          \label{a10}
 \eeq
 \beqa
 W_2&=&\frac{1}{2}\int_{\cal
 D}\left\{-2 \alpha^2\left\lbrack
 (1-\lambda^2)(\bJ\x\bN)\cdot(\bB\cdot\bnab)\bN \right. \right. \nonu
 & & \left. + \lambda(\bN\cdot\bnab\lambda)
 \left(\bN\cdot\frac{\bnab B^2}{2}+N^2\right)\right\rbrack
 \nonu
 && -\alpha\beta\left\lbrack(1-\lambda^2)\bJ\cdot(\bN\x\bM)+2\lambda(\bN\cdot\bnab\lambda)
     \left(\bJ\cdot\frac{\bnab B^2}{2}\right)\right\rbrack \nonu
 && \left. -\alpha\gamma\left\lbrack(1-\lambda^2)\bB\cdot(\bN\x\bM)+2\lambda(\bN\cdot\bnab\lambda)
     \left(\bB\cdot\frac{\bnab B^2}{2}\right)\right\rbrack \right\}
     dV.\nonu
                                                        \label{a11}
 \eeqa
 The coefficients of $\alpha \beta$ and $\alpha\gamma$ vanish
 identically. Indeed, on account of the equilibrium relations
 (\ref{2}), (\ref{5}), (\ref{6}) and (\ref{7}) we have for the coefficient of
 $\alpha\beta$:
 \beqas
 &&(1-\lambda^2)\bJ\cdot(\bN\x\bM)+2\lambda(\bN\cdot\bnab\lambda)
     \left(\bJ\cdot\frac{\bnab
     B^2}{2}\right)\nonu
     && =(1-\lambda^2)g|\bnab\psi|^2\bJ\cdot \bnab
     g+(\lambda^2)^\prime g |\bnab\psi|^2\bJ\cdot\frac{\bnab B^2}{2}
     \nonu
     &&
     =g|\bnab\psi|^2\left\{\bJ\cdot\bnab \left\lbrack
     (1-\lambda^2)g\right\rbrack+\bJ\cdot\bnab\cdot\left\lbrack(\lambda^2)^\prime\frac{B^2}{2}
     \right\rbrack\right\}=0.
 \eeqas
 The coefficient of $\alpha \gamma$ also vanishes because is
 symmetric to the coefficient of $\alpha \beta$ in replacing $\bJ$ with
 $\bB$. Consequently  (\ref{a11}) assumes the form
 (\ref{14}).

 \newpage

 \begin{center}
 {\large\bf Acknowledgements }
 \end{center}

 Part of this work was conducted during a visit of the author G.N.T. to  the
Max-Planck-Institut f\"{u}r Plasmaphysik, Garching. The hospitality
of that Institute is greatly appreciated.

This work was performed  within the participation of the University
of Ioannina in the Association Euratom-Hellenic Republic, which is
supported in part by the European Union and by the General
Secretariat of Research and Technology of Greece. The views and
opinions expressed herein do not necessarily reflect those of the
European Commission.
 \end{document}